\documentclass[10pt]{article}

\usepackage{graphicx,amsmath,amssymb,chemarr}

\newcommand{\beq}{\begin{equation}}
\newcommand{\eeq}{\end{equation}}
\newcommand{\beqn}{\begin{eqnarray}}
\newcommand{\eeqn}{\end{eqnarray}}

\title{Membrane clustering and the role of rebinding in biochemical signaling}

\author{Andrew Mugler$^1$,
	Aimee Gotway Bailey$^2$,
	Koichi Takahashi$^3$,
	Pieter Rein ten Wolde$^{1,}$\thanks{Correspondence: tenwolde@amolf.nl}
	\vspace{.15in} \\
\small $^1$FOM Institute AMOLF, Science Park 104, 1098 XG Amsterdam, The Netherlands\\
\small $^2$American Association for the Advancement of Science, Washington, DC 20005, USA\\
\small $^3$Advanced Sciences Institute, RIKEN, 1-7-22 Suehirocho, Tsurumi, Yokohama, 230-0045, Japan;\\
\small The Molecular Sciences Institute, Berkeley, 2168 Shattuck Avenue, Berkeley, CA 94704;\\
\small Institute for Advanced Biosciences, Keio University, 252-8520 Fujisawa, Japan}

\date{}

\begin{document}

\maketitle

\begin{abstract}
In many cellular signaling pathways, key components form clusters at the cell membrane.  Although much work has focused on the mechanisms behind such cluster formation, the implications for downstream signaling remain poorly understood.  Here, motivated by recent experiments, we study via particle-based simulation a covalent modification network in which the activating component is either clustered or randomly distributed on the membrane.  We find that while clustering reduces the response of a single-modification network, clustering can enhance the response of a double-modification network.  The reduction is a bulk effect: a cluster presents a smaller effective target to a substrate molecule in the bulk.  The enhancement, on the other hand, is a local effect: a cluster promotes the rapid rebinding and second activation of singly active substrate molecules.  As such, the enhancement relies upon frequent collisions on a short timescale, which leads to a diffusion coefficient at which the enhancement is optimal.  We complement simulation with analytic results at both the mean-field and first-passage distribution levels.  Our results emphasize the importance of spatially resolved models, showing that significant effects of spatial correlations persist even in spatially averaged quantities such as response curves.
\end{abstract}

\clearpage

\section*{Introduction}

Although often modeled as well-mixed chemical reactors,
cells are highly spatially heterogeneous entities.
Beyond merely providing the blueprint for space-dependent processes such as division or patterning, spatial heterogeneities in cellular components are frequently exploited by biochemical networks as additional degrees of freedom in signaling computations \cite{Kinkhabwala2010}.  The most direct example is compartmentalization, in which the same chemical component initiates different phenotypic responses depending on where it is localized within the cell \cite{Onken2006, Matallanas2006}.  In a similar way, the localization of signaling components via scaffolding proteins has effects on signal amplification that depend nontrivially on the surrounding chemical conditions \cite{Locasale2007}.
In fact, the colocalization of just two components can have a dramatic response on the amplification properties of a enzyme-driven reaction network \cite{vanAlbada2007}.  Even in spatially uniform systems, spatial correlations between individual molecules can have significant effects on the mean response \cite{Takahashi2010}.

One of the most actively studied areas in which spatial heterogeneity is emerging as a key factor is signal transduction at the cell membrane.  In addition to imposing a quasi-two-dimensional geometry, the membrane plays host to a large diversity of cellular components, the interactions among which give rise to a complex spatial organization \cite{Radhakrishnan2010}.  A central theme of recent work in this field has been the prevalence and role of membrane clusters --- groups of colocalized molecules that often participate in the detection of external signals and subsequently drive responses within the cell.  Perhaps the most well known example occurs in bacterial chemotaxis, in which clusters of receptors detect external ligands, triggering messenger molecules to modulate the activity of flagellar motors \cite{Sourjik2010}.  Evidence for clustering in eukaryotic cell membranes has recently been observed as well: data from immuno-electron microscropy \cite{Plowman2005} and single-molecule fluorescence experiments \cite{Murakoshi2004} suggest that Ras, a protein implicated in a variety of phenotypic responses including oncogenesis, forms membrane clusters on which the efficacy of its downstream signaling critically relies.  Clustering may also be connected to the partitioning of the membrane itself into spatially segmented domains \cite{Goswami2008, vanZanten2009}, e.g.\ via interaction with the cytoskeleton \cite{Machta2011} or formation of so-called {\it lipid rafts} \cite{Simons2010}.

\begin{figure}
   \begin{center}
      \includegraphics*[width=3.25in]{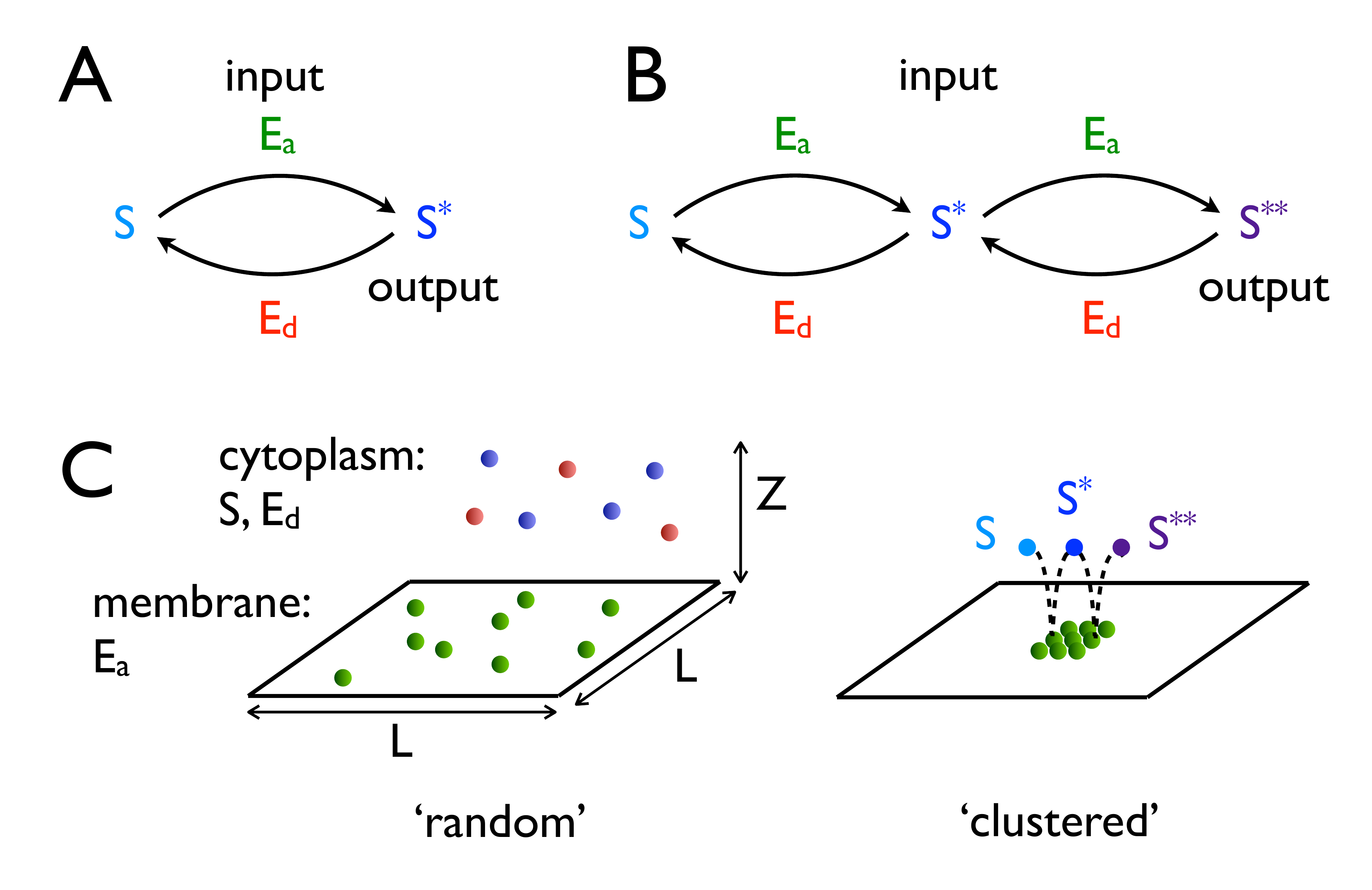}
      \caption{Schematics of reaction networks and spatial arrangement of molecules.  (A) The single-modification network, in which a substrate $S$ is activated and deactivated by components $E_a$ and $E_d$, respectively.  (B) The double-modification network, in which the substrate can be doubly activated.  (C) While $S$ and $E_d$ molecules diffuse freely in the cytoplasm, $E_a$ molecules are fixed on the membrane in either a random (left) or clustered (right) configuration; right panel depicts a rebinding event, in which a singly activated $S^*$ molecule rapidly returns to the $E_a$ cluster to become a doubly activated $S^{**}$ molecule.}
      \label{fig:cartoon}
   \end{center}
\end{figure}

Although much modeling work has been done to elucidate the possible mechanisms by which clusters form \cite{Das2009, Tian2010, Gurry2009}, insights on the role that clustering plays in downstream signaling remain largely speculative.  Therefore a driving goal of the present study is to quantitatively understand, for a spatially resolved model of a canonical signaling network, the effect that clustering has on the input-output response.  Recognizing its ubiquity in the systems in which clustering is observed \cite{Sauro2004}, we focus on a covalent modification network often called the {\it push-pull} network, in which a substrate is alternately activated and deactivated by two antagonistic components (Fig.\ \ref{fig:cartoon}A).  For example, in bacterial chemotaxis, the kinase CheA
and the phosphatase CheZ phosphorylate and dephosphorylate the messenger protein CheY, respectively; CheA and CheZ therefore play the roles of the two antagonistic components, while CheY plays the role of the substrate.

Moreover, focusing on a push-pull network naturally permits extension to a double-modification process (Fig.\ \ref{fig:cartoon}B), which is a critical step in many membrane signaling pathways.  In eukaryotic cells, for example, active Ras molecules at the membrane
initiate a mitogen-activated protein kinase (MAPK) cascade within the cell, each layer of which consists of a dual phosphorylation cycle.  In general, dual phosphorylation can be carried out by one of two mechanisms: in a processive mechanism, an enzyme modifies both phosphorylation sites on a substrate molecule before releasing it; in a distributive mechanism, on the other hand, the enzyme must release the substrate after modification of the first site, before rebinding and modifying the second site.   It has been shown experimentally that key kinases \cite{Burack1997,Ferrell1997} and phosphatases \cite{Zhao2001} in the MAPK cascade act in a distributive manner, making the rebinding process of critical importance.  Therefore a second focus of the present study is to investigate the interplay between clustering and rebinding in determining the input-output response of a distributive push-pull network.

We provide a spatially resolved description of the system by performing particle-based simulations on a lattice.  In parallel, we gain important physical intuition from analytic results derived at both the mean-field and first-passage distribution levels.  We find that the input-output response of the network changes depending on whether the activating component is clustered or randomly distributed on the membrane (Fig.\ \ref{fig:cartoon}C).  Specifically, clustering reduces the response of a single-modification network, while clustering can enhance the response of a double-modification network.  We demonstrate that the reduction is a direct consequence of the fact that a cluster presents a smaller effective target to a substrate molecule in the bulk.  By investigating in detail the stochastic nature of the rebinding process, we discover that the enhancement has an entirely different origin: clustering promotes the rapid rebinding and second activation of singly active substrate molecules (Fig.\ \ref{fig:cartoon}C).  We find that such a rapid effect is only exploited when both the activating and deactivating components are sufficiently free to react, such that {\it ultrasensitive} networks \cite{Goldbeter1981}, in which one or the other component is saturated by the substrate, do not exhibit the enhancement.

What's more, we find that the enhancement with clustering is more pronounced when the diffusion coefficient is large.  Underlying this observation is a fundamental advantage that clustering affords in a collision-dominated regime: while diffusion may be high enough to prevent a substrate molecule from rapidly rebinding an isolated enzyme molecule, it may be insufficient for the substrate molecule to escape an entire cluster.  Clustering thus prolongs the possibility of rapidly rebinding, effectively compensating for low association rates of individual molecules.  Of course, this advantage reaches a limit --- at infinite diffusion all spatial arrangement is forgotten.  We are thus led naturally to a value of the diffusion coefficient at which the enhancement is optimal.

Together our results provide a quantitative picture of the nontrivial effects that membrane clustering has on biochemical signaling, for a network that plays a critical role in systems in which clustering has been experimentally observed.  More broadly, our results demonstrate the crucial role that spatial correlations play in cellular function, and the associated importance of considering spatial resolution in biophysical models.

\section*{Methods}

We consider both a single- and a double-modification push-pull network in which the activating enzyme is localized to the membrane (Fig.\ \ref{fig:cartoon}).
To understand the effect of clustering, we compare the situation in which activating enzyme molecules are arranged randomly on the membrane to that in which they are localized at the same surface density to clusters of size $N$ (Fig.\ \ref{fig:cartoon}C).  Because we are interested in the effect of clustering on downstream signaling, and not in the dynamics of cluster formation on the membrane, we take activating enzyme molecules to be fixed.  Substrate and deactivating enzyme molecules diffuse freely in the cytoplasm with diffusion coefficient $D$.

\subsection*{Chemical reactions, input-output relation, and sensitivity}
The single-modification network (Fig.\ \ref{fig:cartoon}A) is described by the reactions
\beqn
\label{eq:rxn1}
E_a + S \xrightleftharpoons[k_2]{k_1} E_aS \xrightarrow{k_3} E_a + S^*, \\
\label{eq:rxn2}
E_d + S^* \xrightleftharpoons[k_5]{k_4} E_dS^* \xrightarrow{k_6} E_d + S.
\eeqn
Here $S$ and $S^*$ denote the substrate in its inactive and active forms, respectively.  Activation is catalyzed by the activating enzyme $E_a$, which first forms a complex before releasing the substrate in its active state; deactivation is performed similarly by the deactivating enzyme $E_d$.  The double-modification network (Fig.\ \ref{fig:cartoon}B) prescribes additional reactions identical to Eqns.\ \ref{eq:rxn1}-\ref{eq:rxn2}, except with $S$ and $S^*$ replaced by $S^*$ and $S^{**}$, respectively.  We restrict our analysis to networks whose first and second modification processes are identical (i.e.\ the rates $k_1, k_2, \dots, k_6$ describing the first modification also describe the second).  Furthermore all results in this paper assume negligible back reactions: $k_2 = k_5 = 0$.

The {\it input} of a push-pull network is typically defined as either the catalytic rate or the concentration of the activating enzyme.  In the context of signaling via membrane clusters, the former definition finds natural justification: in chemotaxis, for example,
the catalytic activity of the kinase is typically set by the time-averaged ligand occupancy of the receptor cluster \cite{Sourjik2002}.  We therefore take as the input parameter the catalytic rate $k_3$, scaled by its counterpart $k_6$ for the deactivating enzyme: $\chi \equiv k_3/k_6$.  The {\it output} is naturally defined as the relative activity of the substrate, i.e.\ the fraction $\phi \equiv \{ [S^*]/[S]_T, [S^{**}]/[S]_T \}$ for the single- or double-modification network, respectively.  Here $[S]_T$ denotes the total substrate concentration, including all activity states and complexes in which it is involved.  The fraction $\phi$ is a mean quantity, averaged over both time (post-equilibration) and space.  Often we will normalize the output by $\hat{\phi}_{\max}$, the maximum output assuming a deterministic, well-mixed description (Appendix \ref{app:mf}).  The input-output curve $\phi(\chi)$ takes on a characteristically sigmoidal shape (e.g., Fig.\ \ref{fig:N}), becoming particularly sharp, or {\it ultrasensitive}, when the enzymes are saturated by the substrate \cite{Goldbeter1981}.

In a deterministic, well-mixed description, in which rate equations determine the dynamics, the steady-state input-output relation is completely specified by the reaction rates and the conserved total concentrations of substrate $[S]_T$ and enzymes $[E_a]_T$ and $[E_d]_T$.  In particular, for both the single- and the double-modification networks, one may write the input-output relations entirely in terms of the dimensionless parameters (e.g., see Appendix \ref{app:mf})
\beq
\label{eq:dimparams}
\alpha \equiv \frac{[E_d]_T}{[E_a]_T}, \qquad
\beta \equiv \frac{k_4}{k_1}, \qquad
\gamma \equiv \frac{K}{[S]_T}, \qquad
\epsilon \equiv \frac{[E_a]_T}{[S]_T},
\eeq
where $K = k_6/k_4$ is the Michaelis-Menten concentration of the deactivation process, and $[E_a]_T$ is $N$ divided by the volume.
The first two parameters determine the bias of the network toward deactivation; $\alpha=\beta=1$ therefore corresponds to a symmetric network, in which activating and deactivating enzymes have equal concentrations and  association rates to the substrate.  The last two parameters characterize the sensitivity of the network: in the {\it zero-order} (or ultrasensitive) regime, the substrate saturates the enzymes and operates far beyond the Michaelis-Menten concentration ($\{\epsilon, \gamma\} \ll 1$); while in the {\it linear} regime, both substrate and enzymes operate in the linear regions of their response curves ($\{\gamma^{-1}, \epsilon\gamma^{-1}\} \ll 1$).

\subsection*{Spatial lattice model}
Although much can be understood from deterministic, well-mixed descriptions of push-pull networks \cite{Goldbeter1981, Huang1996, Markevich2004}, we expect (and indeed will show) that spatial effects introduced by clustering will significantly influence the signaling properties of the system, even at the level of the mean response.  We therefore perform spatially resolved simulations with excluded volume interactions by introducing a regular three-dimensional lattice.  We make the approximation that all molecules have equal diameter $\ell$, and we let this diameter define the lattice spacing, such that molecules neighboring each other on the lattice are in contact.  Clustered molecules are placed in contact in a square arrangement on the membrane, which is natural given the lattice implementation; we have tested that the results are not significantly affected by instead placing molecules in an arrangement that is circular (up to the lattice resolution).
The membrane comprises the $x$-$y$ plane and extends for a length $L$ in each direction, beyond which periodic boundaries are imposed.  The cytoplasm has depth $Z$, with reflective boundaries at both the membrane ($z=0$) and the farthest point from it ($z=Z$).  Appendix \ref{app:RD} provides a detailed account of how reactions and diffusion are implemented on the lattice, in particular such that detailed balance is obeyed.

Spatial resolution introduces new parameters into the problem beyond those of the well-mixed system (Eqn.\ \ref{eq:dimparams}), which are captured by the following dimensionless quantities.  In addition to the cluster size $N$ one has
\beq
\label{eq:dimparams2}
\delta \equiv \frac{\ell D}{k_1}, \qquad
\mu \equiv \frac{N\ell^2}{L^2}, \qquad
\zeta \equiv \frac{Z}{\ell}.
\eeq
The quantity $1/4\pi\delta = k_1/4\pi\ell D$ is the ratio of the activating enzyme's intrinsic association rate $k_1$, which is the association rate given that the molecules are in contact, to the corresponding diffusion-limited value $4\pi\ell D$; as such $\delta$ captures the strength of diffusion relative to association.  The parameter $\mu$ describes the surface density of activating enzymes and represents the introduction of the lengthscale $L$, beyond which the problem is periodic.  The dimensionless length $\zeta$ reflects the fact that the membrane breaks the $x$-$y$-$z$ symmetry, introducing a second lengthscale $Z$ for the cytoplasmic depth.

Finally, we identify a natural timescale as the time to diffuse approximately one molecular diameter and use it to define a dimensionless time: $\tau \equiv t/(\ell^2/D)$.  Recovery of time-dependent quantities such as first-passage times requires specification of the ratio $\ell^2/D$, while for steady state computations the physical values of the molecular diameter and diffusion coefficient drop out completely; only specification of the dimensionless parameters is required ($\chi$, $N$, and Eqns.\ \ref{eq:dimparams}-\ref{eq:dimparams2}).

\subsection*{Parameter selection}
The Results section discusses in detail the effects of varying the parameters that govern network symmetry ($\alpha$, $\beta$), sensitivity ($\gamma$, $\epsilon$), and diffusion ($\delta$).  In all results establishing network characteristics (Figs.\ \ref{fig:N}, \ref{fig:target}, \ref{fig:capture}, and \ref{fig:SD}), the surface density of activating enzymes ($\mu$) and the cytoplasmic depth ($\zeta$) are set using estimates from experimentally studied systems.  In eukaryotic cells, clustered Ras has been measured to occupy a membrane surface fraction of $\mu \lesssim 1\%$ \cite{Tian2007}. A similar value arises in bacterial chemotaxis: the `long' form of CheA (the form which both associates to receptors and phosphorylates CheY) is present in roughly $4500$ copies per {\it Escherichia coli} cell \cite{Li2004}; taking a cell surface area of $6$ $\mu{\rm m}^2$ and a typical protein diameter of $4$ nm \cite{Phillips2009}, one obtains $\mu\sim 0.012$.  We therefore take $\mu = 0.01$.  The cytoplasmic depth is a measure of the maximum distance from the membrane that a molecule diffuses before encountering a reflective barrier (or, at the most, half the distance to the opposite membrane).  In bacteria, this distance is upper bounded by half the smallest cell lengthscale, or $\sim$$500$ nm.  In eukaryotic cells, this depth is instead dictated by the presence of large organelles near the membrane.  An extreme upper bound can be obtained by noting that organelles comprise roughly half the cell volume \cite{Alberts1994}, and imagining they are spherically packed in the center of a spherical cell of radius $R\sim5$ $\mu$m implies a maximum depth of $Z = [1-(1/2)^{1/3}]R \sim 1000$ nm.  Organelles are, of course, more loosely distributed within the cell, such that a depth on the order of $100$ nm might be more realistic.  We therefore take $Z=100$ nm, which for a molecular diameter of $4$ nm gives $\zeta = 25$.

\section*{Results}

\begin{figure}
\begin{center}
\includegraphics*[width=3.25in]{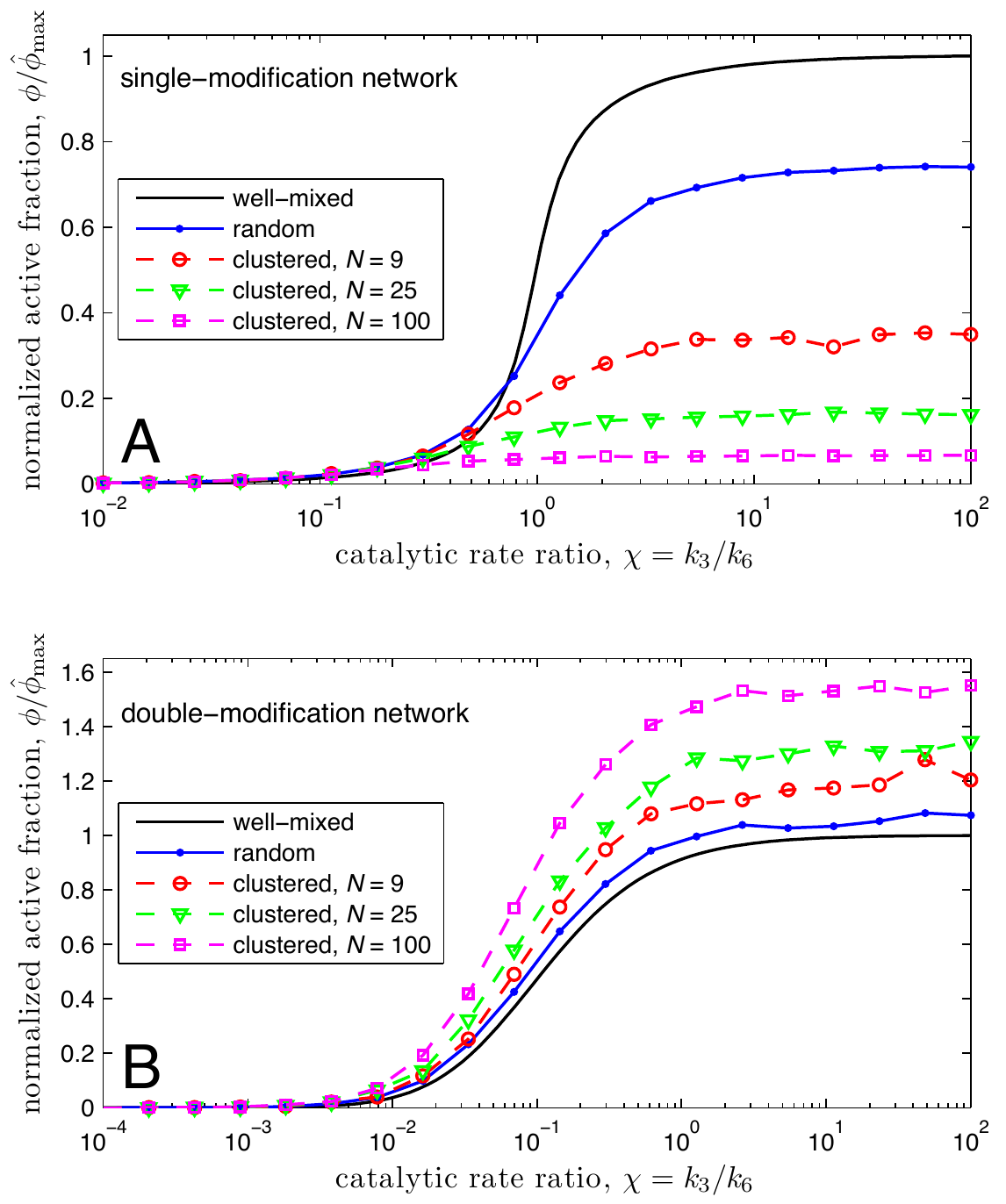}
\caption{Input-output response of a single- and double-modification network.  Cluster size $N$ is varied at constant surface density $\mu=0.01$ and depth $\zeta=25$, leaving the response unchanged with $N$ in the random configuration.  Well-mixed curve is established according to rate equations in steady state (Appendix \ref{app:mf}).  Curves are normalized by the maximal value of the well-mixed response, $\hat{\phi}_{\max}$.  (A) A symmetric ($\alpha=\beta=1$) single-modification network with moderate diffusion ($\delta=1/4\pi$) and zero-order sensitivity ($\gamma=\epsilon=0.1$); here $\phi = [S^*]/[S]_T$.
(B) A deactivation-biased ($\alpha=5$, $\beta=1$) double-modification network with high diffusion ($\delta=10$) and linear sensitivity ($\gamma^{-1}=0.05$, $\epsilon\gamma^{-1}=0.01$); here $\phi = [S^{**}]/[S]_T $.
It is seen that in the clustered configuration, the response is reduced with $N$ for the single-modification network, and enhanced with $N$ for the double-modification network.}
\label{fig:N}
\end{center}
\end{figure}

We begin by presenting and explaining the main difference between the single- and double-modification networks: clustering reduces the response of a single-modification network, while clustering can enhance the response of a double-modification network.  The magnitude of each effect scales with the cluster size $N$ (Fig.\ \ref{fig:N}).  The reduction for single-modification networks is generic, persisting with changes in network symmetry ($\alpha$, $\beta$), sensitivity ($\gamma$, $\epsilon$), and diffusion ($\delta$).  The enhancement for double-modification networks, on the other hand, is more specific, occurring in deactivation-biased linear-sensitivity networks with high diffusion; subsequent results in this section will explain this specificity.

Figure \ref{fig:N} also illustrates more generally the effect of localizing the activating enzymes to the membrane
by comparing the spatially averaged response to the response in the well-mixed case (see Appendix \ref{app:mf}).  As seen in Fig.\ \ref{fig:N}A, localization reduces the maximal response of a single-modification network (compare the `well-mixed' curve to the `random' curve).  Such a reduction was seen in previous work \cite{vanAlbada2007} and is the result of the concentration gradients that form due to the asymmetric localization of activating and deactivating enzymes.  As seen in Fig.\ \ref{fig:N}B, the double-modification network can avoid this reduction and can in fact achieve an amplification beyond the well-mixed response instead.

\subsection*{Clustering reduces the effective target size}

\begin{figure}
\begin{center}
\includegraphics*[width=3.25in]{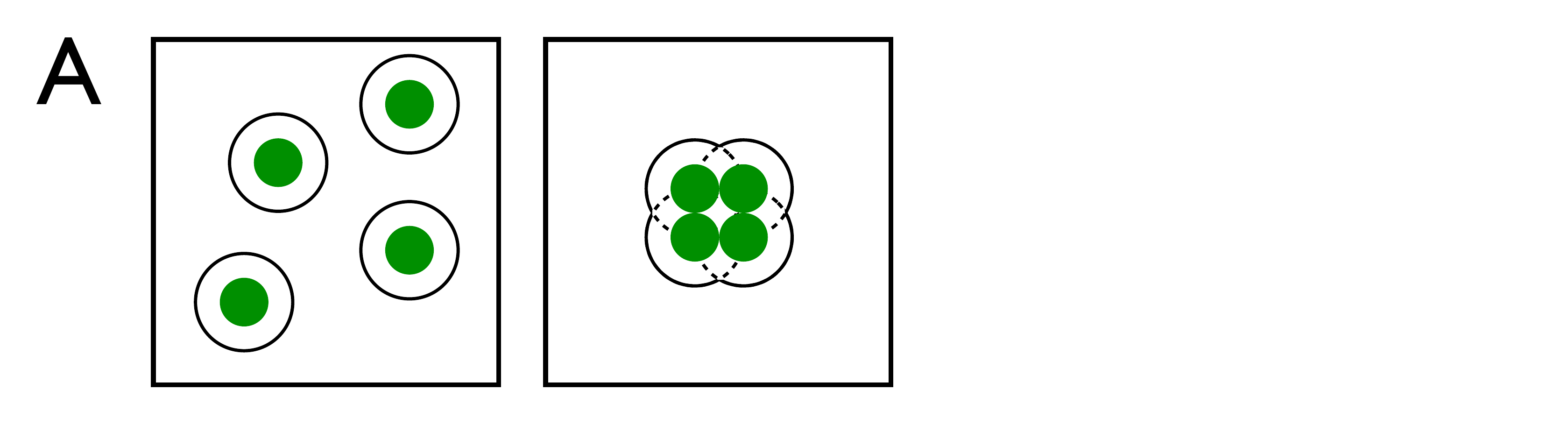}
\includegraphics*[width=3.25in]{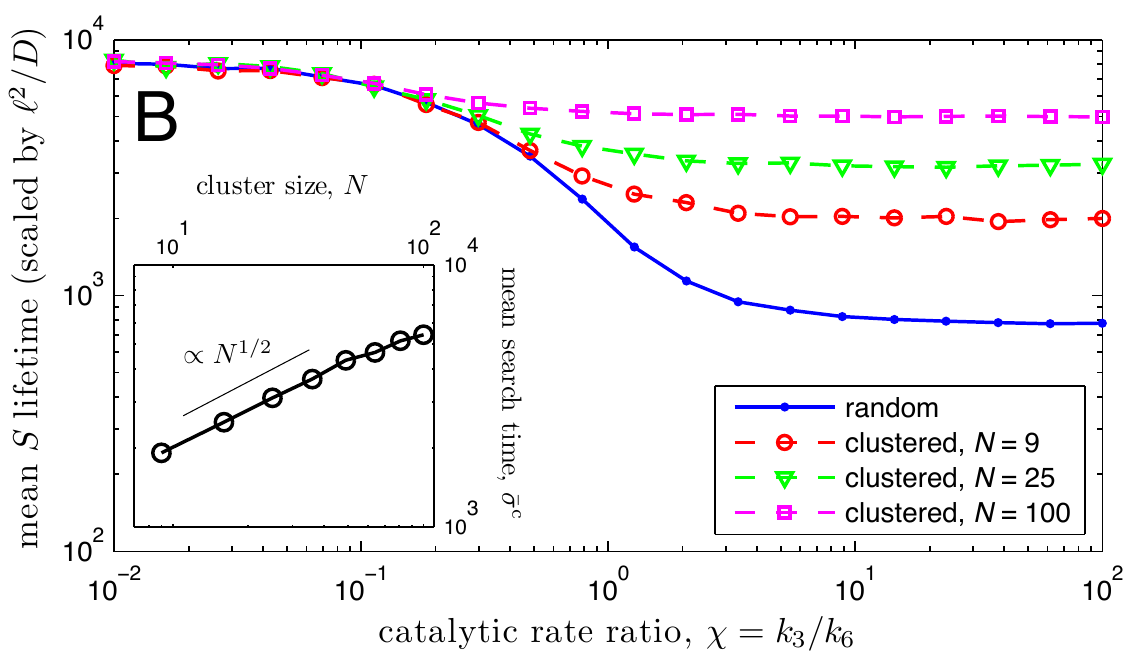}
\caption{The target size effect.  (A) Cartoon depicting the reaction volume boundaries (black lines) of activating enzyme molecules (green circles) in a random (left) and clustered (right) configuration.
The total reaction volume is smaller in the clustered configuration due to the overlap of individual molecules' volumes.
(B) Main plot shows the mean lifetime of $S$ molecules in the single-modification network; parameters are as in Fig.\ \ref{fig:N}A.  Inset demonstrates that the value to which the mean lifetime asymptotes at high input $\chi$ in the clustered configuration --- the mean search time $\bar{\sigma}^{\rm c}$ --- grows as the square root of the cluster size $N$.}
\label{fig:target}
\end{center}
\end{figure}

How does clustering reduce the response of a single-modification network?
The key is that a cluster presents a smaller effective target to a molecule in the bulk.
To understand this fact, one may consider that each molecule possesses a {\it reaction volume} --- the volume that must be entered by the center of mass of a second molecule in order for an association reaction to take place (Fig.\ \ref{fig:target}A).  When the $E_a$ molecules are arranged in a random configuration at low enough density, the total reaction volume (or {\it target size}) is simply $N$ times an individual $E_a$ molecule's reaction volume.  However, when the $E_a$ molecules are clustered, the individual reaction volumes overlap, and the total target size is reduced (Fig.\ \ref{fig:target}A).

To understand quantitatively the impact of the target size reduction on the response of the network, we consider the time it takes an $S$ molecule, released from the bulk, to bind an $E_a$ molecule on the membrane (the {\it lifetime} of the $S$ molecule).  If the $E_a$ molecules are free with high probability (i.e.\ unoccupied by other substrate molecules) the lifetime is dominated by the search time $s$, the time to find and bind an $E_a$ molecule.  The mean search time from the bulk can be estimated as the inverse of the association rate over the volume of the box: $\bar{s} \approx (k_1/L^2Z)^{-1} = L^2Z\delta/\ell D$.  A random distribution of $E_a$ molecules presents $N$ targets of diameter $\ell$, which reduces the mean search time by a factor of $N$: $\bar{s}^{\rm r} = \bar{s}/N = L^2Z\delta/N\ell D$.  A cluster, on the other hand, presents one target with effective diameter $\ell_{\rm eff} \equiv \sqrt{N}\ell$, making the mean search time $\bar{s}^{\rm c} = L^2Z\delta/(\sqrt{N}\ell) D$.  In terms of the dimensionless parameters, these times read $\bar{\sigma}^{\rm r} \equiv
s^{\rm r}/(\ell^2/D) = 
\zeta\delta/\mu$ and $\bar{\sigma}^{\rm c} \equiv
s^{\rm c}/(\ell^2/D) = 
\sqrt{N}\zeta\delta/\mu$, which makes clear that at constant surface density the search time is independent of $N$ for the random configuration but scales with $N^{1/2}$ for the clustered configuration.

Two important conditions of the above analysis are that the $E_a$ molecules are free and that the $S$ molecule is released randomly from the bulk (and not, say, still within the neighborhood of the cluster).  The first condition is met at high input ($\chi \equiv k_3/k_6 \gg 1$), when the high catalytic rate of the activating enzymes leaves the $E_a$ molecules free with high probability.  The second condition is also met at high input for networks with zero-order sensitivity, in which saturation of the deactivating enzymes leaves the $E_d$ molecules occupied with high probability.  High occupation of $E_d$ molecules means that a typical $S^*$ molecule has ample time to randomize its position before ultimately binding a free $E_d$ molecule and being released as an $S$ molecule.  Thus, for $S^*$ molecules the arrangement of the $E_a$ molecules is forgotten, while for $S$ molecules the arrangement is critical.  Indeed, at high input, clustering leads to substrate molecules spending more time in the inactive state, and thus to a reduced output (Fig.\ \ref{fig:N}A).

Fig.\ \ref{fig:target}B shows the mean lifetime of $S$ molecules as a function of the input $\chi$ for a single-modification network with zero-order sensitivity.
At high input, the mean lifetime asymptotes to the value corresponding to the search from the bulk, consistent with the above analysis.  It is clear that for the clustered configuration, this asymptotic value depends on the cluster size $N$, and the inset shows that it indeed scales with $N^{1/2}$, as predicted.

It is important to emphasize that the target size effect is a bulk effect, not a local effect, in the sense that clustering does not only reduce the number of neighboring sites from which a substrate molecule can bind, but more generally reduces the number of distinct {\it paths} which lead to the target from a point in the bulk.  This intuition is confirmed by a simple test: under an alternative implementation, in which a substrate molecule can only bind an $E_a$ molecule from the neighboring lattice site perpendicular to the membrane, we observe an increase in the search time with cluster size $N$ that is only slightly less pronounced than that in the inset of Fig.\ \ref{fig:target}B (not shown).  Because this alternative implementation has the property that clustering the $E_a$ molecules does not change the number of available neighboring sites, the increase in search time with $N$ is strictly due to a reduction in the number of paths from which the target is accessible.

It is also important to point out that the target size effect is a generic property of diffusive random walks, and as such it is just as present for double-modification networks as it is for single-modification networks.  However, as we will describe next, in a particular parameter regime the effects of rapid rebinding can overcome the target size effect, leading to an enhancement of the response rather than a reduction.

Finally, we note that in networks with linear sensitivity, $E_d$ molecules remain free, even at high input (Appendix \ref{app:mf}).  This freedom violates one of the assumptions of the simple scaling analysis above.  Nonetheless, we find that the reduction with clustering in the response of single-modification networks persists.  Next we discuss the role of rebinding in double-modification networks, which will ultimately also give insight into this persistence for single-modification networks.

\subsection*{Clustering promotes rapid rebinding}

How does clustering enhance the response of a double-modification network?
The key is that a cluster promotes rapid rebinding of singly activated substrate molecules.  Rebinding only occurs in the double-modification network; in the single-modification network, once released by an enzyme, a substrate molecule can only bind to an enzyme of the opposite type.  To clearly understand the rapid rebinding effect in double-modification networks, we first consider the distribution of rebinding times for a reduced system: a single $S^*$ molecule is released from one of $N$ $E_a$ molecules, with no $E_d$ molecules present.  It rebinds to any $E_a$ molecule in a time $r$, whose dimensionless analog we define as $\rho \equiv r/(\ell^2/D)$.

\begin{figure}
\begin{center}
\includegraphics*[width=3.25in]{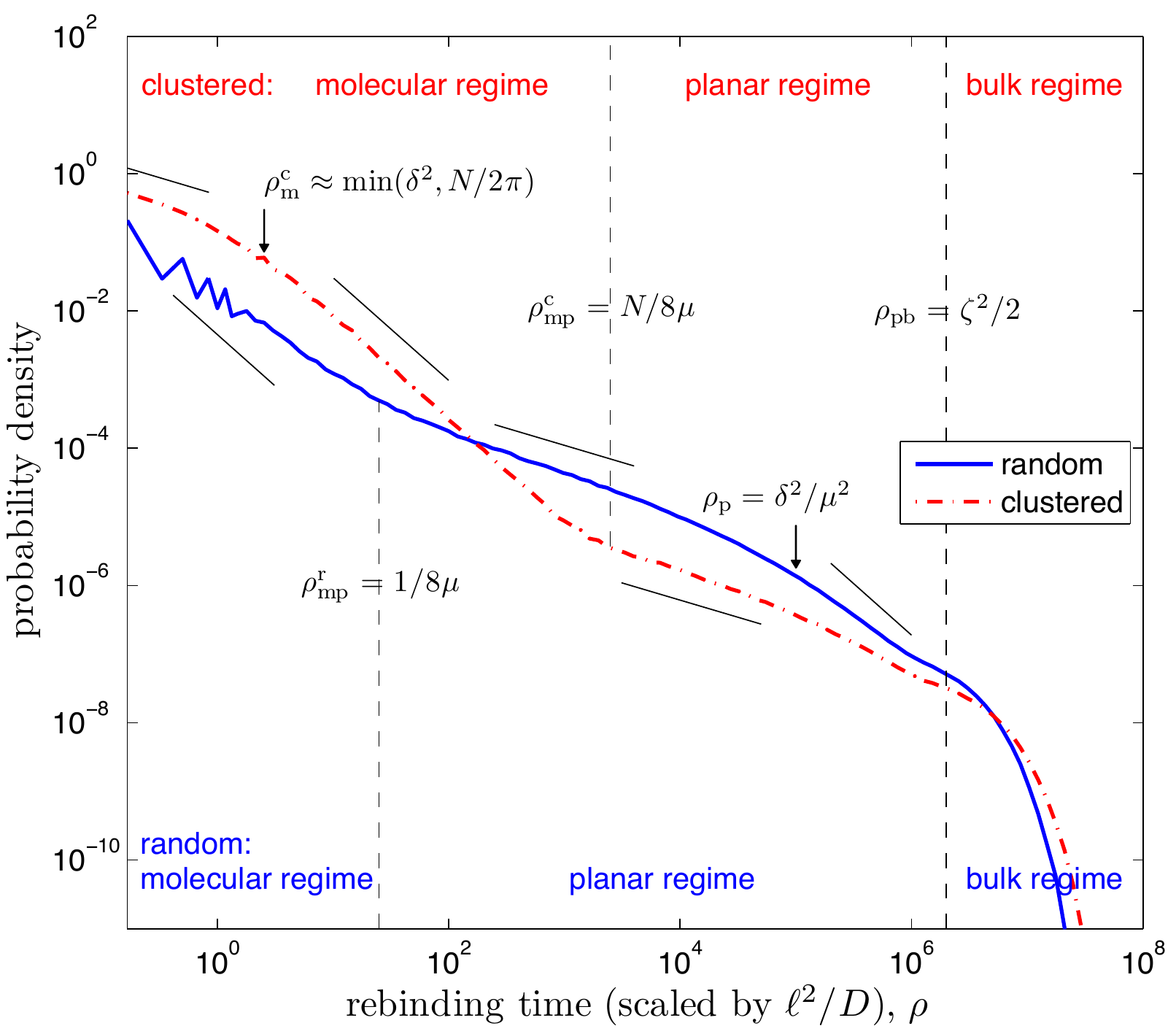}
\caption{Rebinding time distributions.  Distribution of times $\rho$ for a single $S^*$ molecule, released from an $E_a$ molecule, to rebind to an $E_a$ molecule, when the $E_a$ molecules are either randomly distributed (solid) or clustered (dash-dot) on the membrane; no $E_d$ molecules are present.  Short straight lines scale with $\rho^{-1/2}$ (shallow) and $\rho^{-3/2}$ (steep).  To elucidate scalings, high values are taken for the dimensionless parameters describing cluster size ($N=100$), diffusion ($\delta=1.6$), inverse surface density ($\mu^{-1}=200$), and cytoplasmic depth ($\zeta=2000$).  For each of the two configurations, there emerge three regimes distinguishable by scalings.  Times separating the regimes, as well as times marking scaling crossovers within regimes, are derived in the text and indicated on the figure: $\rho_{\rm mp}^{\rm r}$ and $\rho_{\rm mp}^{\rm c}$ separate the molecular from the planar regime in the random and clustered configurations, respectively; $\rho_{\rm bp}$ separates the planar from the bulk regime in both configurations; $\rho_{\rm m}^{\rm c}$ marks the scaling crossover within the molecular regime in the clustered configuration; and $\rho_{\rm p}$ marks the scaling crossover within the planar regime in both configurations.}
\label{fig:rebind}
\end{center}
\end{figure}

As seen in Fig.\ \ref{fig:rebind}, for both a random and clustered configuration of $E_a$ molecules, the rebinding time distribution contains three regimes.  Short times (the {\it molecular} regime) correspond to short excursions, after which the $S^*$ molecule rebinds to the same $E_a$ molecule (or cluster) from which it came.  Intermediate times (the {\it planar} regime) correspond to excursions that are sufficiently far for the $S^*$ to ``see'' the membrane as a plane uniformly populated with $E_a$ molecules (or, due to the periodicity, with clusters), yet not far enough to see the reflective boundary; the granularity of individual $E_a$ molecules is thus lost, and the membrane appears as a uniform semi-absorbent plane.  Long times (the {\it bulk} regime) correspond to long excursions, during which the $S^*$ molecule randomizes its position completely, returning as if from the bulk.

\paragraph{Bulk regime}
The bulk regime exhibits an exponential distribution because it describes the time to find an $E_a$ molecule given a random starting position in the bulk.  An exponential distribution is expected from a well-mixed system, in which reactions obey exponential waiting time statistics.
Here, however, the molecular and planar regimes emerge due entirely to spatial correlations, affecting the network even at the level of the mean response (Fig.\ \ref{fig:N}B).  Moreover, in the bulk regime, the substrate molecule has strayed far enough from the membrane that it effectively returns from the bulk; this return time is therefore equivalent to the search time defined above for $S$ molecules in single-modification networks.  Accordingly, one notices in Fig.\ \ref{fig:rebind} that the time constant characterizing the exponential is larger in the clustered case, precisely due to the target size effect previously discussed.  Finally, the onset of the bulk regime is determined by the time it takes the $S^*$ molecule to randomize its position, which is approximately the time to diffuse the full cytoplasmic depth: $r_{\rm pb} = Z^2/2D$, or $\rho_{\rm pb} = \zeta^2/2$.

\paragraph{Planar regime}
In the planar regime, the substrate molecule has diffused not far enough to enter the bulk so that it loses memory of its starting position, but far enough that the membrane appears as a uniform semi-absorbent plane.  The problem can be reduced to an effectively one-dimensional one in the $z$ direction with a radiation boundary at $z=0$.  The one-dimensional rate $k_{\rm eff}$ (with dimensions of length per time) describing association at the boundary follows from a renormalization of the three-dimensional rate $k_1$; clearly, $k_{\rm eff}$ should scale with the surface density $N/L^2$, and we find good agreement with the simplest dimensionally consistent definition, $k_{\rm eff} \equiv k_1N/L^2$.

As shown in Appendix \ref{app:1D}, the rebinding time distribution for this one-dimensional problem is readily obtained from the Green's function and exhibits scalings of $\rho^{-1/2}$ at short times, $\rho^{-3/2}$ at long times, and a crossover time of $\rho_{\rm p} = (D/\ell k_{\rm eff})^2 = \delta^2/\mu^2$.  Short times comprise a collision-dominated subregime, in which the excursion is dominated by many unsuccessful reflections, and thus inherits the $t^{-1/2}$ scaling from the Gaussian Green's function of a particle freely diffusing in one dimension.  Long times comprise a search-dominated subregime, in which after a long excursion the particle returns to an effectively absorbing boundary, producing the $t^{-3/2}$ scaling characteristic of a one-dimensional random walker returning to an absorbing origin.  Further detail is provided in Appendix \ref{app:1D}.

The transition between the molecular and planar regimes occurs when the $S^*$ molecule diffuses far enough perpendicular to the membrane that it no longer detects the granularity of the $E_a$ molecules, a distance roughly equal to half the mean spacing between $E_a$ molecules in the random configuration, or between clusters in the clustered configuration.  In the random configuration, the mean spacing between $E_a$ molecules is set by the surface density, yielding a separating time of $r_{\rm mp}^{\rm r} = [\sqrt{(L^2/N)}/2]^2/2D$, or $\rho_{\rm mp}^{\rm r} = 1/8\mu$.  In the clustered configuration, the spacing between clusters is $L$, yielding a separating time of $r_{\rm mp}^{\rm c} = (L/2)^2/2D$, or $\rho_{\rm mp}^{\rm c} = N/8\mu$.

\paragraph{Molecular regime}
The molecular regime is defined by short excursions, in which the substrate molecule rebinds to the $E_a$ molecule or cluster from which it came.  The molecular regime exhibits $\rho^{-1/2}$ and $\rho^{-3/2}$ scalings whose origins are the same as those in the planar regime: the scalings arise from a collision-dominated or search-dominated return, respectively, to a single molecule or cluster.  For a return to single molecule, which applies to the random configuration, these scalings were described in previous work \cite{Takahashi2010}.  The crossover time was derived to be
\beq
\label{eq:rm}
r_{\rm m} = \frac{\ell^2/D}{(1+k/4\pi\ell D)^2},
\eeq
where here $k = 2k_1$, the factor of two arising from reflection of the $E_a$ across the membrane.
One sees from Fig.\ \ref{fig:rebind}, however, that in the random configuration the crossover time is obscured by alternations in the probability density at short times, which is an artifact of the lattice implementation.  To be precise: an $S^*$ molecule starting next to an $E_a$ molecule can only rebind in an odd number of time steps (assuming it moves diffusively every time step, which is true at short times for large $\delta$); the exception occurs when another $E_a$ molecule is placed next to or very near the first $E_a$ molecule, but at low surface densities such a placement occurs with low probability.  We have validated the distributions in Fig.\ \ref{fig:rebind} using Green's Function Reaction Dynamics \cite{Takahashi2010}, verifying that lattice artifacts do not quantitatively change the probability densities.

In the clustered configuration, the crossover time within the molecular regime is indeed resolvable and can be described in terms of the previously considered results.  A large, absorbent cluster ($N\gg 1$, $\delta \ll 1$) can be approximated as a plane with an effective one-dimensional association rate $k_{\rm eff} \equiv k_1/\ell^2$, yielding a dimensionless crossover time of $(D/\ell k_{\rm eff})^2 = \delta^2$.  In the opposite limit, a small, reflective cluster ($N\sim 1$, $\delta \gg 1$) can be approximated as a spherical object whose effective diameter is obtained by equating surface areas:
$4\pi (\ell_{\rm eff}/2)^2 = 2N\ell^2$ (neglecting cluster edges).
In the limit of large $\delta$ the denominator in
Eqn.\ \ref{eq:rm}
approaches unity, making the crossover time approximately $\ell_{\rm eff}^2/D$, or $N/2\pi$ in dimensionless units.  Since the expressions in both the plane and sphere limits scale with parameters that are large in the opposite limits, we use the minimum as an estimate of the crossover time: $\rho_{\rm m}^{\rm c} \approx \min(\delta^2, N/2\pi)$.
\\
\\
Figure \ref{fig:rebind} corroborates all scalings and crossover times derived above using an illustrative set of sample parameters.
Because we have analytic estimates for the crossover times, they can be tuned to expand or contract the various regimes, a fact we have used to confirm the validity of the scalings beyond the confidence implied by Fig.\ \ref{fig:rebind} alone.

Figure \ref{fig:rebind} also directly displays the advantage that clustering affords in the rebinding problem: at short times, the probability of rebinding is enhanced, leading to a {\it probability gap} over the random configuration.  In fact, the characteristic time that determines the extent of this gap, $\rho_{\rm m}^{\rm c}$, reveals the parameter regimes that give rise to enhanced rebinding, and thus ultimately to an enhanced signal output for the network.  Specifically, the gap increases as $\rho_{\rm m}^{\rm c}$ increases, both via increasing the cluster size $N$ and increasing diffusion relative to association, $\delta$.  Increasing the cluster size is a straightforward way of enhancing rebinding, and the associated enhancement of the output is demonstrated in Fig.\ \ref{fig:N}B.

The reason that increasing diffusion increases the probability gap is perhaps less straightforward but can be understood at the molecular level.  High diffusion induces many unsuccessful collisions before eventual rebinding.  Rapidly rebinding to a single $E_a$ molecule (which is the task when the $E_a$ molecules are randomly distributed) is is therefore unlikely.  Rapidly rebinding to a cluster, on the other hand, is less unlikely, owing to the presence of neighbors.  The number of collisions in the neighborhood of a cluster is simply larger than that for a single molecule by virtue of its increased size.  The probability of ultimately achieving a successful collision is thus higher for the clustered configuration than for the random configuration, by a factor that increases as the system is placed more strongly in the diffusive, or collision-dominated, regime.

At a more detailed mechanistic level, we may consider the fate of an $S^*$ molecule that has just been released by an $E_a$ molecule, and now resides at a neighboring lattice point.  At high diffusion, the probability is large that the $S^*$ molecule will take a step away from the $E_a$ molecule.  In the random configuration, it is then increasingly likely for diffusion to carry the $S^*$ molecule away from the immediate vicinity of the $E_a$ molecule.  In the clustered configuration, on the other hand, several of these diffusive paths will lead directly to another $E_a$ molecule.  Clustering therefore poses an advantage when high diffusion ensures that immediate rebinding is unlikely, but rebinding after several diffusive steps is more probable.

Of course, the advantage afforded by clustering cannot persist in the infinite diffusion (well-mixed) limit --- in this limit, all spatial information is lost.  The consequences of this fact for the network output are discussed in more detail later in this section.

\subsection*{Deactivation connects rebinding to the network response}

Interestingly, despite the probability gap elucidated above, we observe that the means of the two rebinding distributions in Fig.\ \ref{fig:rebind} are the same: the enhancement conferred to the clustered configuration in the molecular regime is compensated by the target size effect in the bulk regime.  The equivalence of means is a consequence of the fact that we have isolated the rebinding process.  Alone, the rebinding process is equivalent to one dissociation and subsequent association event of the equilibrium reaction $A + B \rightleftharpoons C$.  For equilibrium reactions, detailed balance ensures that mean quantities are unaffected by spatial arrangement.  Therefore, although the push-pull network as a whole prescribes a non-equilibrium process, the rebinding process alone is effectively in equilibrium, and the mean rebinding time is the same for a random and a clustered configuration.

How then does the probability gap translate to an enhancement with clustering at the level of the mean response, as in Fig.\ \ref{fig:N}B?  Indeed, it is precisely because thus far in the discussion we have not reintroduced the $E_d$ molecules.  The effect of the $E_d$ molecules is to bind and deactivate the $S^*$ molecules with the longest excursion times, removing them from the rebinding problem and thereby truncating the rebinding distributions beyond a characteristic timescale, which we call the {\it capture} time.  The truncation alleviates the target size effect, imparting the clustered configuration with a shorter mean rebinding time than the random configuration.

\begin{figure}
\begin{center}
\includegraphics*[width=3.25in]{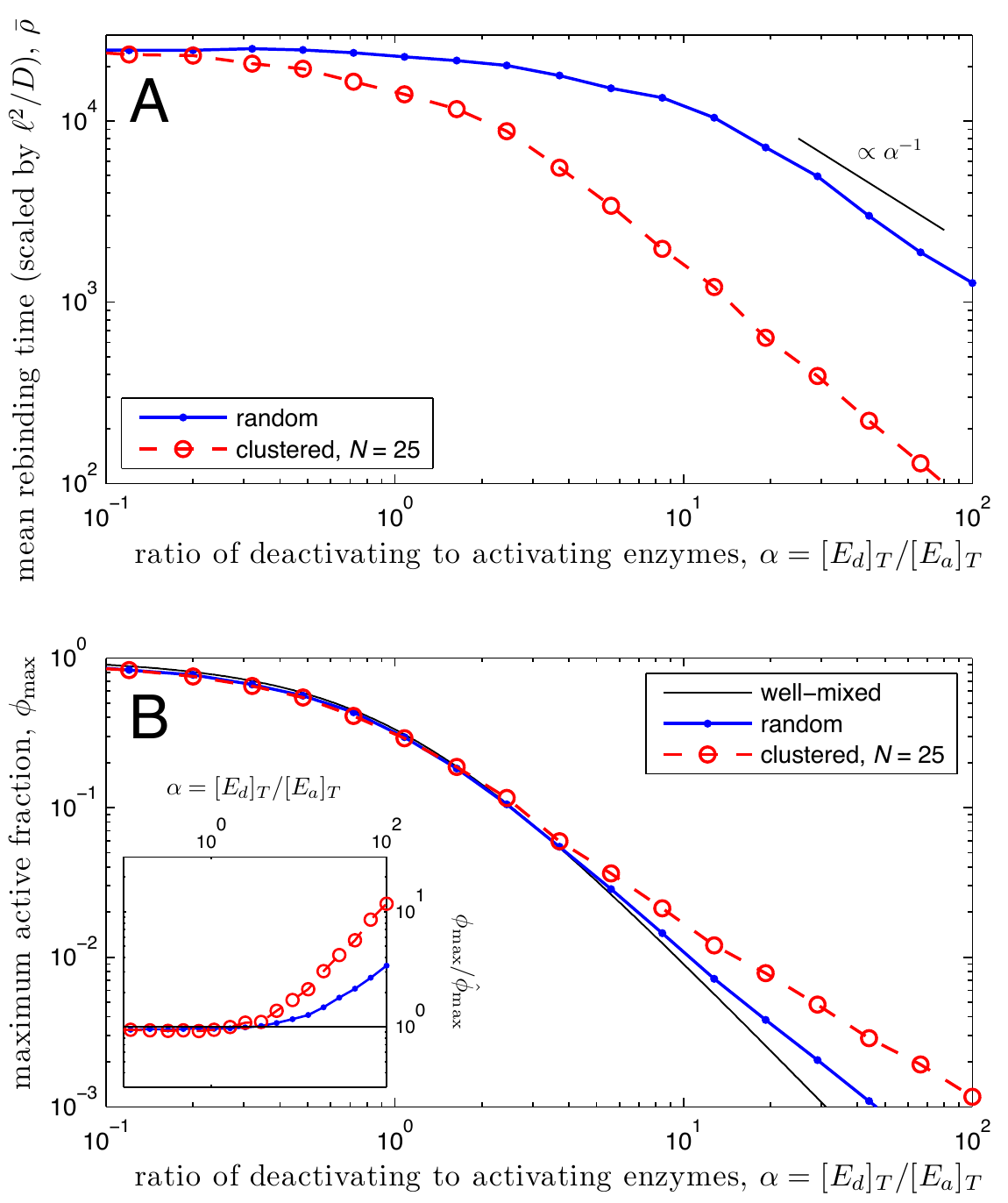}
\caption{The capture effect.  (A) As deactivating enzymes are introduced, increasing the enzyme ratio $\alpha$, the mean rebinding time $\bar{\rho}$ decreases more quickly for a clustered configuration than for a random configuration.  Here $\mu=0.01$, $\zeta=25$, $N=25$, and $\delta=10$.  (B) Correspondingly, in a double-modification network, the maximal output $\phi_{\max}$ for the clustered configuration is higher than that for the random configuration at large $\alpha$.  Solid line shows the analytic prediction in the well-mixed limit, from which the data deviate due to spatial effects.  Inset shows same data normalized by the well-mixed prediction $\hat{\phi}_{\max}$.  Sensitivity is linear ($\gamma^{-1}=0.05$, $\epsilon\gamma^{-1}=0.01$) with  $\beta=1$ and other parameters as in A.}
\label{fig:capture}
\end{center}
\end{figure}

The capture effect is illustrated in Fig.\ \ref{fig:capture}A, in which $E_d$ molecules are gradually reintroduced into the system.  With $E_d$ molecules present, an $S^*$ molecule has two fates $f$: it may rebind to an $E_a$ molecule ($f=+$) or be captured by an $E_d$ molecule ($f=-$); measuring the time $\tau$ for either fate samples the joint distribution $p(\tau,f)$.  The mean rebinding time $\bar{\rho}$ is then computed from the conditional distribution $p(\tau|+) = p(\tau,+)/p(+)$, where $p(+)$ is the total probability of rebinding as opposed to capture.  Figure\ \ref{fig:capture}A shows that the difference in mean rebinding times between the random and clustered configurations, $\Delta\bar{\rho} \equiv \bar{\rho}^{\rm r}-\bar{\rho}^{\rm c}$, indeed increases as the ratio $\alpha$ of $E_d$ to $E_a$ molecules is increased.  The increase abates at large $\alpha$, when the truncation at the capture time dominates the rebinding distribution, such that $\bar{\rho}$ approaches the capture time.  The capture time can be estimated as the inverse of the association rate times the concentration of $E_d$ molecules, $(k_4[E_d]_T)^{-1}$, or $\delta\zeta/\alpha\beta\mu$ in dimensionless units; accordingly, Fig.\ \ref{fig:capture}A demonstrates that $\bar{\rho}$ scales with $\alpha^{-1}$ at large $\alpha$.

The result of the capture effect for the double-modification network is illustrated in Fig.\ \ref{fig:capture}B, which shows the maximal output $\phi_{\max}$ as a function of $\alpha$ for a network with linear sensitivity.  One observes that the clustered configuration produces a larger $\phi_{\max}$ than the random configuration beyond an enzyme ratio of $\alpha^* \sim 1.5$ (see inset), indicating that clustering enhances the output when there are more deactivating enzymes than activating enzymes, such that the capture effect is strong.
Indeed, at $\alpha^*$ the difference in mean rebinding times ($\Delta\bar{\rho} \sim 9\times 10^3$)
sufficiently surpasses the corresponding difference in mean search times
($\Delta\bar{\sigma} \equiv \bar{\sigma}^{\rm c}-\bar{\sigma}^{\rm r} \approx 3\times 10^3$) --- that is, the capture effect overtakes the target size effect.

Lastly, we point out that the rebinding distributions in Fig.\ \ref{fig:rebind}, although strictly a component of double-modification networks, also lend intuition to the analysis of single-modification networks.  Although rebinding does not occur directly in single-modification networks, the process of an $S^*$ being released from an $E_a$, binding an $E_d$, being released as an $S$, and returning to an $E_a$ can be thought of as an effective rebinding excursion.  The fact that deactivation must occur during this excursion imposes a minimum rebinding time, effectively introducing a truncation of the rebinding distributions at {\it short} times, thereby imparting the clustered configuration with a {\it longer} mean rebinding time than the random configuration.  This result helps explain why the reduction in output seen with clustering in single-modification networks (Fig.\ \ref{fig:N}A) persists even in networks with linear sensitivity: since the $E_d$ molecules are free with high probability even at high input, the mean deactivation time is the same for both configurations; the extra time in the clustered configuration must therefore be spent while the substrate is inactive, leading to a lower output on average.

\subsection*{Clustering benefits double-modification networks with linear sensitivity}

Figure \ref{fig:capture}B clearly illustrates a trade-off: at high $\alpha$, clustering enhances the network output beyond that of the random configuration, but increasing $\alpha$ reduces the maximal output in general.
The reduction with $\alpha$ can be derived in the deterministic, well-mixed limit (Appendix \ref{app:mf}); the result, $\hat{\phi}_{\max} = (1+\alpha\beta+\alpha^2\beta^2)^{-1}$, is overlaid in Figure \ref{fig:capture}B and provides a baseline from which the data diverge due to spatial effects.
The intuition behind the reduction is straightforward: biasing the network toward deactivation reduces the fraction of active substrate.  The reduction, however, is unique to networks with linear sensitivity, which leads to the question: can clustering enhance the network response in the zero-order regime, in which the maximal output remains high?

\begin{figure}
\begin{center}
\includegraphics*[width=3.25in]{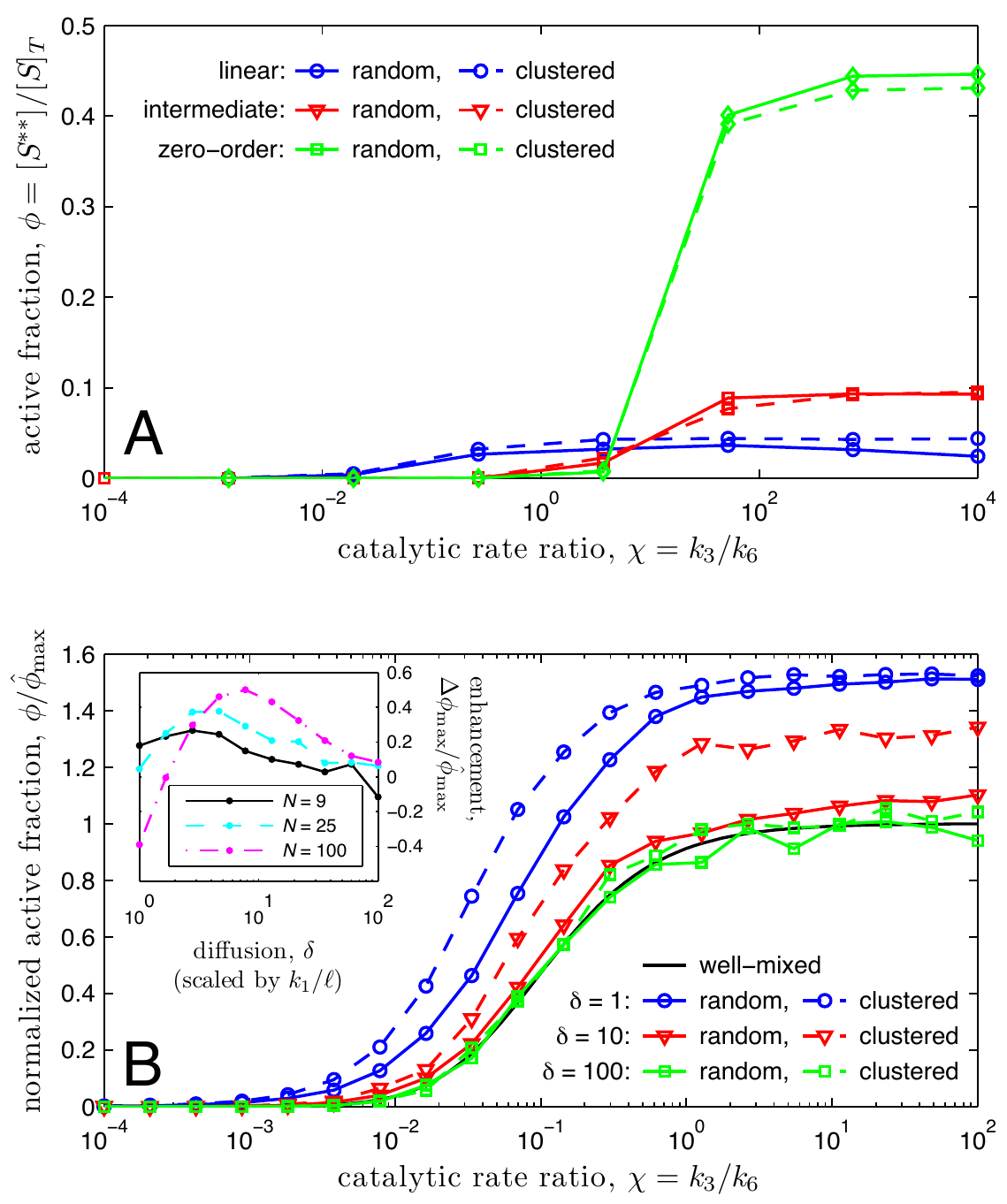}
\caption{The effects of varying sensitivity and diffusion on the response of a double-modification network.  Both panels take $N=25$, $\alpha=5$, $\beta=1$, $\mu = 0.01$, and $\zeta=25$. (A) The input-output response of a network with high diffusion ($\delta=10$) as sensitivity is varied from linear ($\gamma^{-1}=0.05$, $\epsilon\gamma^{-1}=0.01$) to intermediate ($\gamma=0.2$, $\epsilon=0.1$) to zero-order ($\gamma=0.05$, $\epsilon=0.05$). (B) The input-output response of a linear network ($\gamma^{-1}=0.05$, $\epsilon\gamma^{-1}=0.01$) as diffusion is varied.  The inset shows the normalized difference between the maximal output in the clustered and random configurations (the {\it enhancement}) versus $\delta$ for several values of the cluster size $N$.}
\label{fig:SD}
\end{center}
\end{figure}

The answer, revealed in Fig.\ \ref{fig:SD}A, is no: as the sensitivity is varied from linear to zero-order, the enhancement with clustering vanishes, then reverses.  The reason is that the capture effect, which underlies the enhancement with clustering, relies on the $E_d$ molecules being free.  The zero-order regime corresponds to saturation of the enzymes by the substrate, such that at high input the $E_d$ molecules are not free, but rather they are occupied with high probability.  The mean capture time is then exceedingly long, such that the mean rebinding time is independent of the configuration of $E_a$ moelcules.  The target size effect takes over, and the random configuration produces a higher output.
The end result is that the benefit of clustering is specific to double-modification networks with linear sensitivity; in ultrasensitive networks the benefit is lost.

\subsection*{Clustering leads to an optimal diffusion coefficient}

In discussing the rebinding distributions, it was discovered that increasing diffusion enhances rapid rebinding to a cluster more strongly than to a random configuration, because it places the system in a more collision-dominated regime.  One then expects the enhancement with clustering to increase with the diffusion coefficient.  However, we also know that at very high diffusion, the network is well-mixed, and the spatial arrangement of the molecules is irrelevant; clustering should therefore confer no enhancement at very high diffusion.  In fact, these competing effects lead to a diffusion coefficient at which the enhancement is optimal, as shown in Fig.\ \ref{fig:SD}B.

Figure \ref{fig:SD}B illustrates that as the ratio $\delta$ of diffusion to association is increased, the enhancement, i.e.\ the difference in maximal output between the clustered and the random configuration $\Delta \phi_{\max} \equiv \phi_{\max}^{\rm c} - \phi_{\max}^{\rm r}$, first increases then decreases.  The inset shows this nonmonotonic behavior for several values of the cluster size $N$.
The optimal enhancement increases with $N$; moreover, the value $\delta^*$ at which the optimum occurs also increases with $N$.  These observations are consistent with the notion that a larger cluster can confer an advantage more effectively in a highly diffusive regime.

Quantitatively, $\delta^*$ approaches $\sim$$10$ for the largest cluster size ($N=100$), corresponding to an association rate $k_1 \sim \ell D/10$, roughly $10\cdot 4\pi \approx 120$ times less than the diffusion-limited value.  Optimal enhancement therefore occurs in a regime in which unsuccessful collisions are very frequent because association is very rare.  This result suggests a possible function of clustering in a double-modification system, in terms of compensating for weak association rates of individual enzymes: by promoting rapid rebinding events, a cluster enhances the total probability of associating.  This function may have evolved as a way for a cell to boost the response when intrinsic association rates are very low.

\section*{Discussion}

We provide a detailed view of the varied effects that membrane clustering can have on the signaling properties of a canonical biochemical network.  The network under study and the values of relevant biophysical parameters are drawn from experimentally studied systems, both prokaryotic and eukaryotic, in which membrane clustering has been recently observed.  We implement a spatially resolved model, appealing to both simulation and analytic results to demonstrate that spatial correlations can have nontrivial effects, even at the level of the mean input-output response.  In particular, we show that spatial effects at both the bulk scale --- in terms of a diffusive target search process --- and at the molecular scale --- in terms of rapid stochastic rebinding events --- affect the response of a network in ways that are not captured by a well-mixed, spatially uniform description.

Our results make clear that the effect of clustering depends on both the network topology and the biochemical parameters.  For example, we identify a general property of diffusive random walks --- that clustering the target increases the search time from the bulk --- which leads generically to a reduced response in a single-modification network.  However, when the topology of the network is extended to double-modification, the reduction can be overcome by a local effect: clustering promotes rapid rebinding of singly active substrate molecules to the activating enzyme molecules.  When the concentration of free deactivating enzyme molecules is sufficiently high to isolate these rapid rebinding events, the result is an enhancement of the response.  Importantly, this enhancement is specific to networks with linear sensitivity; ultrasensitive networks, in which the deactivating enzyme molecules are saturated by the substrate at high input, do not confer the enhancement because the mechanism relies on the deactivating enzyme molecules being free.  Moreover, the enhancement is most pronounced in a highly diffusive regime, in which unsuccessful collisions dominate, making the probability to escape a single activating enzyme molecule much higher than that to escape a cluster.  The specificity of the enhancement to both linear sensitivity and high diffusion highlights the importance of biochemical parameters in predicting the effects of clustering.

The result that clustering is most beneficial in a highly diffusive regime has important functional implications.  We find that the diffusion coefficient at which the cluster-induced enhancement is optimal corresponds to an intrinsic association rate roughly $100$ times smaller than its diffusion-limited value.  This finding implies that clustering is most helpful for signaling when intrinsic association rates are very low.  Mechanistically, such a result is sensible, since the presence of clustered neighbors can boost the overall probability of rebinding, even when the probability of rebinding to a single molecule is negligible.  Functionally, this result suggests that membrane clustering may have evolved as a way to boost signal output despite low intrinsic association rates.  Indeed, such a result could be tested experimentally: our study predicts that enzymes that are involved in clustered signaling complexes are likely to have intrinsic association rates that are lower than the diffusion-limited value, placing the system in a collision-dominated regime.  It would be interesting to test this prediction with a controlled assay of binding affinites or a curation of binding rates for proteins that are known to cluster.

Our findings emphasize the important role of the rebinding process in biochemical signaling.  The importance of rebinding has been discussed in related systems, with interesting consequences for the mean response.  For example, in studying how the diffusive motion of a repressor protein effects gene expression, it has been observed that the mean response remains describable by a well-mixed theory, albeit with parameters that are appropriately rescaled to account for enhanced noise \cite{VanZon2006}.  In studying signaling via a MAPK cascade, on the other hand, it has been found that spatial correlations due to rapid rebinding introduce qualitative changes in the mean response that cannot be captured by the well-mixed theory \cite{Takahashi2010}.  Our results here are more resonant with the second case, since it is clear that membrane localization and subsequent clustering introduce changes to the rebinding statistics (Fig.\ \ref{fig:rebind}) that go beyond the exponential distributions expected from a well-mixed description.  More broadly, the importance of rebinding has been recognized in explaining the potency of T cell ligand binding, in which a long aggregated binding time arises from a sequence of many fast rebinding events \cite{Govern2010}.

This study represents a first step in using simulation and analytic techniques to understand the role of spatial organization in signaling.  It is our view that spatially resolved models, as well as a sharp theoretical framework, can help formalize and make more quantitative the inferences that are being made from the wealth of experimental data on systems which exhibit clustering, colocalization, and other nontrivial spatial heterogeneity.

\section*{Acknowledgments}
We thank Nils Becker for useful discussions, and Christopher Govern for a critical reading of the manuscript.
This work is part of the research program of the ``Stichting voor Fundamenteel 
Onderzoek der Materie (FOM)'', which is financially supported by the ``Nederlandse 
organisatie voor Wetenschappelijk Onderzoek (NWO)''.

\appendix

\section{Reaction-diffusion implementation on the lattice}
\label{app:RD}

In this appendix, we describe how reactions and diffusion are implemented for particles on the lattice.  In particular, the implementation obeys detailed balance and ensures that the deterministic results (next section) are recovered by spatially averaged quantities in the high-diffusion limit.

We remind the reader that we consider a regular three-dimensional lattice with excluded volume interactions.  We make the approximation that all molecules have equal diameter $\ell$, and we let this diameter define the lattice spacing, such that molecules neighboring each other on the lattice are in contact.  In the clustered configuration, $N$ molecules are placed in contact in a square arrangement on the membrane.
The membrane comprises the $x$-$y$ plane and extends for a length $L$ in each direction, beyond which periodic boundaries are imposed.  The cytoplasm has depth $Z$, with reflective boundaries at both the membrane ($z=0$) and the farthest point from it ($z=Z$).  Reflection at $z=0$ implies that cytoplasmic molecules do not bind directly to the membrane, but rather only bind to the cytoplasmic domain of membrane-bound molecules.

Particle numbers used in the simulation box of volume $L^2Z$ are expressible in terms of the dimensionless parameters in Eqns.\ \ref{eq:dimparams}-\ref{eq:dimparams2}: the number of activating enzyme molecules is $N$, the number of deactivating enzyme molecules is $\alpha N$, and the number of substrate molecules is $N/\epsilon$.

Over timescales longer than the time to diffuse a few molecular diameters, rotational diffusion sufficiently randomizes a molecule's orientation.  Thus although we imagine each deactivating enzyme as possessing a catalytic domain to which the substrate binds, we model its reaction propensities as isotropically distributed over its surface.  Since the activating enzymes, on the other hand, are membrane-bound, the situation is more subtle: we suppose that the cytoplasmic domain of each activating enzyme molecule traces out the hemispherical solid angle inside the membrane, except when blocked by neighboring activating enzymes (a consideration particularly relevant when the activating enzymes are clustered).
Neighbors thus have the effect of reducing the molecule's cross-section:
the reaction propensity of each activating enzyme molecule is distributed over the portion of its surface both inside the membrane and unblocked by neighbors.

We note that the system as described can be mapped to a statistically equivalent system with periodic boundaries in the $z$ direction, which offers both simpler implementation and, in some cases, more direct analytic interpretation.  Specifically, the reflective boundaries at $z=0$ and $Z$ are equivalent to a periodic boundary at $z=0$ and $2Z$, so long as we recognize that the cytoplasmic molecules then double in number and the membrane-bound molecules (having been reflected across the membrane) become twice as reactive.
The periodic boundaries confer the advantage that the membrane no longer needs to be explicitly implemented in the simulation: cytoplasmic molecules can occupy the plane in which the activating enzyme molecules reside, and substrate molecules can bind to activating enzyme molecules from any free neighboring site.

We describe the implementation of reactions and diffusion on the lattice using a simple example,
then extend to the push-pull reactions
(Eqns.\ \ref{eq:rxn1}-\ref{eq:rxn2}).
We consider the binary reversible reaction $A+B\rightleftharpoons C$, in which association and dissociation are described by intrinsic rates $k_a$ (with dimensions of length cubed per time) and $k_d$ (with dimensions of inverse time), respectively.  Dissociation is modeled as a first-order reaction event with an exponential waiting time distribution; the probability for a $C$ molecule to dissociate in a time step $d t$ is thus $p_d \approx k_dd t$ for small $p_d$.  Association is set by detailed balance, which equates the ratio of microscopic probabilities to enter and leave a reaction state to the ratio of macroscopic rates \cite{Morelli2008}; on a lattice with spacing $\ell$, the detailed balance condition reads $p_a\ell^3/p_d = k_a/k_d$, yielding $p_a = (k_a/\ell^3)d t$ for the association probability of an $A$ and $B$ molecule at contact.
Finally, diffusion is implemented according to its microscopic definition, namely
that the mean squared distance traveled in a time $dt$ is $6Dd t$; the six possible moves on the lattice result a mean squared distance of $6p_D\ell^2$ in one time step, making $p_D = (D/\ell^2)d t$ the probability for a molecule to diffuse to a neighboring site.

We now extend the above expressions to the push-pull reactions and write them in terms of the dimensionless parameters (Eqns.\ \ref{eq:dimparams}-\ref{eq:dimparams2}), the cluster size $N$, the input $\chi \equiv k_3/k_6$, and the dimensionless time $\tau \equiv t/(\ell^2/D)$.  The probability to diffuse to a neighboring site in a time step $d t$ is $p_D = (D/\ell^2)d t = d\tau$.  There are two association reactions, with probabilities of occurring from contact $p_1 = (k_1/\ell^3)d t = (1/\delta)d\tau$ (activation) and $p_4 = (k_4/\ell^3)d t = (\beta/\delta)d\tau$ (deactivation).  Lastly, there are two dissociation reactions, with probabilities of occuring $p_3 = k_3 d t = (\chi\beta\gamma\mu/\delta\epsilon\zeta)d\tau$ (activation) and $p_6 = k_6 d t = (\beta\gamma\mu/\delta\epsilon\zeta)d\tau$ (deactivation), where in addition to using the definitions of the dimensionless parameters, we have recognized explicitly that $[E_a]_T = N/L^2Z$.
The time step $d\tau$ is chosen small enough that the sum of each molecule's diffusion and reaction probabilities is bounded from above by one at all times.

Both association and dissociation probabilities are divided uniformly over the faces of each molecule (or, in the case of membrane-bound molecules, the faces not blocked by fixed neighbors).
Furthermore, the choice of which molecule actually moves during a dissociation event is determined by diffusion: in the binary reaction, $A$ would move with probability $D_A/(D_A+D_B)$ and $B$ would move with probability $D_B/(D_A+D_B)$.  In practice, then, when a substrate molecule dissociates from an activating enzyme molecule, the substrate molecule always moves, because the activating enzymes are fixed.  When a substrate molecule dissociates from a deactivating enzyme molecule, on the other hand, each molecule moves with probability $1/2$, because the diffusion coefficients are equal.  These choices ensure that the total probability of associating or dissociating in each time step sums to $p_a$ or $p_d$, respectively, and therefore that detailed balance is obeyed.

\section{Analytic results in the deterministic, well-mixed limit}
\label{app:mf}

In this appendix, we derive key analytic results for both the single- and double-modification push-pull network in the deterministic, well-mixed limit (i.e.\ invoking rate equations).  Strictly speaking, these results are exact for averaged quantities in the limit of infinite diffusion.  More broadly, however, they lend powerful intuition to the spatially resolved results, even when diffusion plays a significant role.

\subsection{The single-modification network}

We begin with the single-modification network (Eqns.\ \ref{eq:rxn1}-\ref{eq:rxn2}), which in steady state is described by the rate equations
\beqn
\label{eq:rate1}
0 &=& \frac{d[S]}{dt} = -k_1[E_a][S] + k_6[E_dS^*], \\
\label{eq:rate2}
0 &=& \frac{d[S^*]}{dt} = -k_4[E_d][S^*] + k_3[E_aS], \\
\label{eq:rate3}
0 &=& \frac{d[E_a]}{dt} = -\frac{d[E_aS]}{dt} = -k_1[E_a][S] + k_3[E_aS], \\
\label{eq:rate4}
0 &=& \frac{d[E_d]}{dt} = -\frac{d[E_dS^*]}{dt} = -k_4[E_d][S^*] + k_6[E_dS^*].
\eeqn
Here, as in the main text, we neglect back reactions ($k_2=k_5=0$).  The rate equations are complemented by the conservation laws
\beqn
[E_a]_T &=& [E_a] + [E_aS], \\
{[}E_d]_T &=& [E_d] + [E_dS^*], \\
\label{eq:con3}
[S]_T &=& [S] + [S^*] + [E_aS] + [E_dS^*].
\eeqn
As implied by the conservation laws, the rate equations contain three redundancies from the zero net flux of activating enzyme, deactivating enzyme, and substrate; two are made explicit in Eqns.\ \ref{eq:rate3}-\ref{eq:rate4}, and the third is revealed by the fact that the sum of Eqns.\ \ref{eq:rate1}-\ref{eq:rate2} equals the sum of Eqns.\ \ref{eq:rate3}-\ref{eq:rate4}.   Eqns.\ \ref{eq:rate1}-\ref{eq:con3} thus constitute six independent equations for six unknowns.  Scaling concentrations by the Michaelis-Menten concentration of the deactivation process, $K\equiv k_6/k_4$,
\beq
\label{eq:dimvars}
x_1 \equiv \frac{[E_a]}{K}, \qquad
x_2 \equiv \frac{[S]}{K}, \qquad
x_3 \equiv \frac{[E_d]}{K}, \qquad
x_4 \equiv \frac{[E_aS]}{K}, \qquad
x_5 \equiv \frac{[E_dS^*]}{K},
\eeq
and recalling the definitions of the dimensionless parameters introduced in Eqn.\ \ref{eq:dimparams},
\beq
\label{eq:dimparamsapp}
\alpha \equiv \frac{[E_d]_T}{[E_a]_T}, \qquad
\beta \equiv \frac{k_4}{k_1}, \qquad
\gamma \equiv \frac{K}{[S]_T}, \qquad
\epsilon \equiv \frac{[E_a]_T}{[S]_T},
\eeq
the six independent equations may be written
\beqn
\label{eq:single1}
x_1x_2 &=& \beta x_5, \\
\label{eq:single2}
\chi^{-1}x_3\phi &=& \gamma x_4, \\
x_3\phi &=& \gamma x_5, \\
\epsilon &=& \gamma(x_1+x_4), \\
\alpha\epsilon &=& \gamma(x_3+x_5), \\
\label{eq:single6}
1 &=& \phi+\gamma(x_2+x_4+x_5),
\eeqn
where $\chi \equiv k_3/k_6$ and $\phi \equiv [S^*]/[S]_T$ are the input and output, respectively.

Combining Eqns.\ \ref{eq:single1}-\ref{eq:single6} yields a third-degree polynomial equation for $\phi$ \cite{Goldbeter1981}:
\beqn
0 &=& \chi(\chi-\alpha)\phi^3
	+ \left[ \alpha(1+\chi)(\chi-\alpha)\epsilon + \chi(2\chi+\alpha\beta\chi-\alpha)\gamma
		- \chi(\chi-\alpha) \right]\phi^2 \nonumber \\
\label{eq:exact}
&&	+ \chi \left[ \alpha(1+\chi)\epsilon + \chi(1+\alpha\beta)\gamma - (2\chi-\alpha) \right] \gamma \phi
	- \chi^2 \gamma^2.
\eeqn
In principle, Eqn.\ \ref{eq:exact} can be solved for the input-output relation $\phi(\chi)$.  However, the solution to such a cubic equation is quite unwieldy, and we therefore focus on limits of Eqn.\ \ref{eq:exact} (or the original Eqns.\ \ref{eq:single1}-\ref{eq:single6}) at high input ($\chi \gg 1$), and in the zero-order ($\{\epsilon, \gamma\} \ll 1$) and linear regimes ($\{\eta, \nu\} \ll 1$).  Here, for notational convenience, we have defined
\beq
\label{eq:etanu}
\eta \equiv \frac{1}{\gamma} = \frac{[S]_T}{K}, \qquad
\nu \equiv \frac{\epsilon}{\gamma} = \frac{[E_a]_T}{K}.
\eeq

We find the maximal output value, $\phi(\chi \gg 1) \equiv \phi_{\max}$, directly from Eqns.\ \ref{eq:single1}-\ref{eq:single6}.  The leading order result is obtained when $\chi^{-1}=0$ exactly; by Eqn.\ \ref{eq:single2} this leads to $x_4=0 \Rightarrow [E_aS]=0$, which makes sense because at infinite catalytic rate $k_3$ the complex $E_aS$ has zero lifetime.  Combining the five remaining equations yields a quadratic equation,
\beq
\label{eq:xmaxeq}
0 = \phi_{\max}^2 + [\alpha\epsilon + (1+\alpha\beta)\gamma-1]\phi_{\max} -\gamma,
\eeq
whose solution directly gives $\phi_{\max}$.

In the zero-order regime, to zeroth order in the small parameters ($\gamma=\epsilon=0$), Eqn.\ \ref{eq:xmaxeq} reads $0=\phi_{\max}(\phi_{\max}-1)$, giving the maximal output value $\phi_{\max}=1$: it is possible to activate all substrate molecules at high input.  Further insight is revealed by Eqn.\ \ref{eq:exact}, which for $\gamma=\epsilon=0$ reduces to
\beq
\label{eq:switch}
0 = (\chi-\alpha)\phi^2(\phi-1).
\eeq
Here, when $\chi \ne \alpha$, $\phi$ must be either $0$ or $1$, implying switch-like behavior around the threshold $\chi^* = \alpha$.  This switch-like behavior
is the hallmark of zero-order sensitivity \cite{Goldbeter1981}.

In the linear regime, we obtain $\phi_{\max}$ by rewriting Eqn.\ \ref{eq:xmaxeq} in terms of $\eta$ and $\nu$:
\beq
\label{eq:xmaxeq2}
0 = \eta \phi_{\max}^2 + [\alpha\nu + (1+\alpha\beta)-\eta]\phi_{\max} -1.
\eeq
To zeroth order in the small parameters ($\eta=\nu=0$), Eqn.\ \ref{eq:xmaxeq2} gives
\beq
\label{eq:xmax2}
\phi_{\max} = \frac{1}{1+\alpha\beta}.
\eeq
Interestingly, for a symmetric network ($\beta=\alpha=1$), we see that in the linear regime it is only possible to activate half the substrate molecules at high input.  We also obtain the threshold value from Eqn.\ \ref{eq:exact}, which in terms of $\eta$ and $\nu$ reads
\beqn
0 &=& \chi(\chi-\alpha)\eta^2\phi^3
	+ \left[ \alpha(1+\chi)(\chi-\alpha)\nu + \chi(2\chi+\alpha\beta\chi-\alpha)
		- \chi(\chi-\alpha)\eta \right]\eta\phi^2 \nonumber \\
\label{eq:exact2}
&&	+ \chi\left[ \alpha(1+\chi)\nu + \chi(1+\alpha\beta) - (2\chi-\alpha)\eta \right]\phi
	- \chi^2.
\eeqn
Taking Eqn.\ \ref{eq:exact2} to first order in $\eta$ and $\nu$ yields
\beq
0 = \eta(2\chi+\alpha\beta\chi-\alpha)\phi^2
	+ \left[ \alpha(1+\chi)\nu + \chi(1+\alpha\beta) - (2\chi-\alpha)\eta \right]\phi
	- \chi,
\eeq
into which we insert $\phi = \phi_{\max}/2 = 1/[2(1+\alpha\beta)]$ and solve for $\chi$, yielding the threshold value
\beq
\label{eq:linthresh}
\chi^* = \frac{\alpha(1+2\alpha\beta)\eta + 2\alpha(1+\alpha\beta)\nu}
	{2(1+\alpha\beta)^2 + (1+3\alpha\beta)\eta - 2\alpha(1+\alpha\beta)\nu}
	\approx \left[ \frac{\alpha(1+2\alpha\beta)}{2(1+\alpha\beta)^2} \right] \eta
	+ \left[ \frac{\alpha}{1+\alpha\beta} \right] \nu.
\eeq
We see that while in the zero-order regime the threshold is set by the ratio of activating to deactivating enzymes, $\alpha$, in the linear regime the threshold vanishes in proportion to the small parameters that define the regime, $\eta$ and $\nu$.  We find that Eqn.\ \ref{eq:linthresh} also serves as a good estimate for the threshold in a double-modification network with linear sensitivity, and thus explains why the threshold shifts to small $\chi$ in
Fig.\ \ref{fig:SD}A of the main text.

\subsection{The double-modification network}

We now consider the double-modification network, which prescribes additional reactions identical to Eqns.\ \ref{eq:rxn1}-\ref{eq:rxn2}, except with $S$ and $S^*$ replaced by $S^*$ and $S^{**}$, respectively.  As in the main text, we restrict our analysis to networks whose first and second modification processes are identical (i.e.\ the rates $k_1, k_2, \dots, k_6$ describing the first modification also describe the second).  There are now nine species, described by the dimensionless variables in Eqn.\ \ref{eq:dimvars}, the new variables
\beq
\label{eq:dimvars2}
x_6 \equiv \frac{[S^*]}{K}, \qquad
x_7 \equiv \frac{[E_aS^*]}{K}, \qquad
x_8 \equiv \frac{[E_dS^{**}]}{K},
\eeq
and the redefined output $\phi \equiv [S^{**}]/[S]_T$.  As before the nine rate equations contain three redundancies from the zero net flux of activating enzyme, deactivating enzyme, and substrate; together with the three conservation laws we thus have nine independent equations for nine unknowns:
\beqn
\label{eq:double1}
x_1x_2 &=& \beta x_5, \\
\label{eq:double2}
\chi^{-1} x_6x_3 &=& x_4, \\
x_6x_3 &=& x_5, \\
x_1x_6 &=& \beta x_8, \\
\label{eq:double5}
\chi^{-1} \phi x_3 &=& \gamma x_7, \\
\phi x_3 &=& \gamma x_8, \\
\epsilon &=& \gamma(x_1+x_4+x_7), \\
\alpha\epsilon &=& \gamma(x_3+x_5+x_8), \\
\label{eq:double9}
1 &=& \phi+\gamma(x_2+x_4+x_5+x_6+x_7+x_8).
\eeqn

Although it is no longer straightforward to combine Eqns.\ \ref{eq:double1}-\ref{eq:double9} into a single polynomial in $\phi$, results in certain limits can be obtained directly from the equations themselves.  For example, we may immediately seek the maximal output value $\phi_{\max}$ by taking the limit $\chi \gg 1$ to zeroth order ($\chi^{-1} = 0$).  By Eqns.\ \ref{eq:double2} and \ref{eq:double5} this limit leads to $x_4=0 \Rightarrow [E_aS]=0$ and $x_7=0 \Rightarrow [E_aS^*]=0$, respectively, which make sense because at infinite catalytic rate $k_3$ the complexes $E_aS$ and $E_aS^*$ have zero lifetime.  Although the remaining seven equations still do not lead easily to a single equation for $\phi_{\max}$, it is possible to derive an expression for $\phi_{\max}$ directly in the linear regime.

The linear regime implies that the parameters $\eta$ and $\nu$ are small (Eqn.\ \ref{eq:etanu}); in terms of these parameters, the remaining seven equations read:
\beqn
\label{eq:sat1}
x_1x_2 &=& \beta x_5, \\
\label{eq:sat2}
x_6x_3 &=& x_5, \\
\label{eq:sat3}
x_1x_6 &=& \beta x_8, \\
\label{eq:sat4}
\eta \phi_{\max} x_3 &=& x_8, \\
\nu &=& x_1, \\
\alpha\nu &=& x_3+x_5+x_8, \\
\label{eq:sat7}
\eta &=& \eta \phi_{\max}+x_2+x_5+x_6+x_8.
\eeqn
Furthermore, since the linear regime is defined by the fact that both substrate and enzyme concentrations are much smaller than $K$, the dimensionless variables $x_i$ are also small (Eqns.\ \ref{eq:dimvars}, \ref{eq:dimvars2}).  We may then recognize that Eqns.\ \ref{eq:sat1}-\ref{eq:sat2} and \ref{eq:sat3}-\ref{eq:sat4} imply that $x_5$ and $x_8$, respectively, are small to second order.  To first order, then, $x_5=x_8=0$, and
\beqn
\label{eq:satlin1}
x_1x_2 &=& \beta x_6x_3, \\
\label{eq:satlin2}
x_1x_6 &=& \beta\eta \phi_{\max} x_3, \\
\nu &=& x_1, \\
\alpha\nu &=& x_3, \\
\label{eq:satlin5}
\eta &=& \eta\phi_{\max}+x_2+x_6,
\eeqn
where Eqns.\ \ref{eq:satlin1} and \ref{eq:satlin2} come from combining Eqns.\ \ref{eq:sat1}-\ref{eq:sat2} and \ref{eq:sat3}-\ref{eq:sat4}, respectively.  It is now trivial to solve Eqns.\ \ref{eq:satlin1}-\ref{eq:satlin5} for $ \phi_{\max}$, yielding
\beq
\phi_{\max} = \frac{1}{1+\alpha\beta+\alpha^2\beta^2}.
\eeq
Upon comparing this expression with that for the single-modification network (Eqn.\ \ref{eq:xmax2}), we see that the maximal output for the double-modification network is suppressed by an additional term $\alpha^2\beta^2$ in the denominator.  Indeed, with respect to the deactivating enzymes, if the activating enzymes are fewer ($\alpha>1$) or associate more weakly to the substrate ($\beta>1$), suppression of the output is severe in the linear regime.

Additionally, the result that $x_5$ and $x_8$ are small to second order implies that $[E_dS^*]$ and $[E_dS^{**}]$ are much smaller than $[E_d]$.  To leading order, then, $[E_d]_T\approx [E_d]$, i.e.\ in the linear regime the deactivating enzymes are approximately all free, {\it even at maximal input}.  Indeed, this feature is a primary difference between the two sensitivity regimes: in the zero-order regime either the activating or deactivating enzymes are saturated by the substrate, while in the linear regime both activating and deactivating enzymes are free.  The implications of this freedom for signaling are discussed at several points in the main text.

\section{Rebinding time distribution in one dimension}
\label{app:1D}

In this appendix, we consider the problem of a particle diffusing in a one-dimensional space that is free on one side and has a radiation boundary condition on the other.  The distribution of first-passage times at the boundary is directly obtainable from the Green's function for this problem, which is known.  The result provides a good approximation for the distribution of rebinding times to a planar membrane populated with absorbing constituents, for excursions long enough that the plane appears as a uniform, semi-absorbent sink.

We consider a particle diffusing along the positive $z$-axis with a radiation boundary at $z=0$.  The diffusion equation describes the evolution of the probability density $p(z|t)$:
\beq
\label{eq:DE}
\frac{\partial p(z|t)}{\partial t} = D \frac{\partial^2 p(z|t)}{\partial z^2}.
\eeq
The radiation boundary condition states that the flux leaving the boundary is due to a reaction, which requires both that the particle is at the boundary, with probability $p(0|t)$, and that the reaction fires, with intrinsic rate $k$ (dimensions length per time):
\beq
\label{eq:BC}
D \left. \frac{\partial p(z|t)}{\partial z}\right|_{z=0} = k p(0|t).
\eeq
The solution given that the particle starts at point $z_0$, i.e.
\beq
\label{eq:IC}
p(z|0) = \delta(z-z_0),
\eeq
which is called the Green's function, is known \cite{Beck1992}:
\beqn
p(z|t,z_0) &=& \frac{1}{\sqrt{4\pi Dt}} \left[ e^{-(z-z_0)^2/(4Dt)} + e^{-(z+z_0)^2/(4Dt)} \right] \nonumber\\
\label{eq:GF}
&& -\frac{k}{D} e^{k^2t/D} e^{k(z+z_0)/D} {\rm erfc}\left( \frac{z+z_0}{\sqrt{4Dt}} + \sqrt{\frac{k^2t}{D}} \right).
\eeqn
The first term is the solution for a reflecting boundary, and the second term describes the loss of probability incurred by the reaction; erfc denotes the complementary error function.

The distribution of first-passage times through the boundary $P(t|z_0)$ is equal to the flux out of the boundary at time $t$, which by Eqn.\ \ref{eq:BC} is
\beq
P(t|z_0) = k p(0|t,z_0).
\eeq
We obtain $p(0|t,z_0)$ directly from Eqn.\ \ref{eq:GF}, yielding
\beq
\label{eq:FP}
P(t|z_0) = \frac{k}{\sqrt{\pi Dt}} e^{-z_0^2/(4Dt)}
	- \frac{k^2}{D} e^{k^2t/D} e^{kz_0/D}
	{\rm erfc}\left( \frac{z_0}{\sqrt{4Dt}} + \sqrt{\frac{k^2t}{D}} \right).
\eeq
We specialize to the distribution of {\it rebinding} times $r \equiv t$ by demanding that the particle starts at the boundary, $z_0 = 0$.  Eqn.\ \ref{eq:FP} then becomes
\beqn
P(r) &=& \frac{k}{\sqrt{\pi Dr}} - \frac{k^2}{D} e^{k^2r/D} {\rm erfc}\left(\sqrt{\frac{k^2r}{D}} \right) \\
\label{eq:rebind}
&=& \frac{1}{r_{\rm p}} \left[ \frac{1}{\sqrt{\pi r/r_{\rm p}}}
	- e^{r/r_{\rm p}} {\rm erfc}\left(\sqrt{r/r_{\rm p}}\right) \right].
\eeqn
In the second line we have recognized that reaction and diffusion define a characteristic timescale $r_{\rm p} \equiv D/k^2$.  The meaning of this timescale is elucidated by considering times much shorter or longer than $r_{\rm p}$, as described below.

At short times ($r \ll r_{\rm p}$), the second term in Eqn.\ \ref{eq:rebind} is unity to leading order, and the first term dominates, producing a $r^{-1/2}$ scaling:
\beq
\label{eq:rshort}
P(r) \approx \frac{1}{\sqrt{\pi r_{\rm p}}} r^{-1/2} \qquad r \ll r_{\rm p}.
\eeq
Such short times correspond to a collision-dominated regime: the particle does not diffuse appreciably far from the boundary; instead, it makes quick bounces against the boundary, getting reflected until ultimately becoming absorbed.

The above intuition can be sharpened in two ways.  First, we may quantify the notion of ``appreciably far'' by realizing that reaction and diffusion define a characteristic length $d \equiv D/k$.  The speed at which a particle travels this length in time $r_{\rm p}$ is $d/r_{\rm p} = k$, revealing that the reaction rate $k$ may also be interpreted as the mean velocity at which particles are ``pulled'' into the boundary.  Particles diffusing farther than $d$ escape this radiative pull, while particles remaining within $d$ stay close to the boundary until absorbed.

Second, we may appeal to Bayes's rule to understand the scaling in Eqn.\ \ref{eq:rshort}.  Supposing that a short excursion is comprised of a number of unsuccessful reflections, ultimate absorption requires the radiation reaction to fire {\it given} that the particle is at the boundary ($z=0$).  The probability of this event occurring at time $r$ can be written using Bayes's rule as
\beq
p(r|z=0) = \frac{p(z=0|r)p(r)}{p(z=0)} \propto p(z=0|r)p(r),
\eeq
where $p(z=0) = \int_0^\infty dr\, p(z=0|r)p(r)$ normalizes the distribution and is independent of time.  The first term on the right, for a reflecting particle, is equivalent to the free-particle solution in one dimension: $p(z=0|r) = (4\pi Dr)^{-1/2} e^{-(0)^2/4Dr} \propto r^{-1/2}$.  The second term is described by an exponential waiting time distribution, whose time constant must be given by the only timescale in the problem, $r_{\rm p}$: $p(r) = r_{\rm p}^{-1} e^{-r/r_{\rm p}}$.  For $r \ll r_{\rm p}$, $p(r) \approx r_{\rm p}^{-1}$ is constant to leading order, and $p(r|z=0) \propto r^{-1/2}$, as in Eqn.\ \ref{eq:rshort}.

At long times ($r \gg r_{\rm p}$), the erfc in Eqn.\ \ref{eq:rebind} can be approximated by its asymptotic limit, yielding
\beqn
P(r) &\approx& \frac{1}{r_{\rm p}} \left[ \frac{1}{\sqrt{\pi r/r_{\rm p}}} - e^{r/r_{\rm p}}
	\frac{e^{-r/r_{\rm p}}}{\sqrt{\pi r/r_{\rm p}}}
	\left( 1+ \sum_{n=1}^\infty (-1)^n \frac{1\cdot 3\cdot 5 \cdot \ldots (2n-1)}{(2r/r_{\rm p})^n} \right) \right] \\
&=& \sqrt{\frac{r_{\rm p}}{4\pi}} r^{-3/2} \qquad r \gg r_{\rm p},
\eeqn
to leading order.  Such long times correspond to a search-dominated regime: the time spent far from the boundary diffusing is much greater than the time spent close to the boundary making short reflections.  The process is therefore well approximated by a one-dimensional random walker returning to an absorbing origin, which scales as $r^{-3/2}$.  Indeed, explicitly imposing an absorbing boundary by taking $k\rightarrow \infty$ makes the crossover time $r_{\rm p} \rightarrow 0$, such that the distribution scales as $r^{-3/2}$ for all times.

In the main text, we rescale $r$ by the characteristic time to diffuse a molecular diameter $\ell$, yielding the dimensionless rebinding time $\rho \equiv r/(\ell^2/D)$ and the associated crossover time $\rho_{\rm p} = r_p/(\ell^2/D) = (D/\ell k)^2$.  The $\rho^{-1/2}$ and $\rho^{-3/2}$ scalings, as well as the crossover time $\rho_{\rm p}$, are observed in Fig.\ \ref{fig:rebind} of the main text in the {\it planar} regime, in which a substrate molecule diffuses far enough from the membrane that the problem can be approximated as one-dimensional, but not so far that it encounters the reflective boundary opposite the membrane.


\end{document}